# ABSTRACT

Statistical tests were conducted on 1,000 numbers generated from the Genome of Bacteriophage T4, obtained from GenBank with accession number AF158101. The numbers passed the non-parametric, distribution-free tests. Deoxyribonucleic acid (DNA) was discovered to be a random number generator, existing in nature.

# **Testing for Sub-cellular Randomness**.

**Okunoye Babatunde O.**
*Department of Pure and Applied Biology, Ladoke Akintola University of Technology,
P.M.B. 4000 Ogbomoso Nigeria.  +234 – 0802 – 985 – 4850.*
babatundeokunoye@yahoo.co.uk

# INTRODUCTION

Molecular computations (Adleman, 1994; Quyang *et al,* 1997; Pirrung *et al*, 2000; Sakamoto et al, 2000; Wang et al, 2001) and devices (Benenson et al, 2001; Stojanovic *et al,* 2002; Benenson et al, 2003, Stojanovic &Stefanovic, 2003; Stojanovic & Stefanovic, 2003, Margolin & Stojanovic, 2005) are being constructed with increasing ease, establishing the field of DNA computing.

Computer and biomolecular systems both start from a small set of elementary Components from which, layer by layer, more complex entities are constructed with ever more sophisticated functions (Regev & Shapiro, 2002).

Our accumulated knowledge about biomolecular computing has not been matched by an understanding, in computer science terms, of the principles which govern them. Computer science can provide the much needed abstraction for biomolecular systems (Regev & Shapiro, 2002).

The hallmark of scientific understanding is the reduction of a natural phenomenon to simpler units. An abstraction a mapping from a real-world domain to a mathematical domain highlights essential properties while ignoring other, complicating ones (Regev & Shapiro, 2002).

In pursuit of abstractions therefore, we may wonder why in Leonard Adleman's pioneering and groundbreaking molecular computation (Adleman, 1994), random algorithmic paths were produced. Random algorithmic paths point to randomized algorithms, the existence of which is made possible by the injection of randomness into the behaviour of the algorithm (Karp, 1986). This is usually achieved by connecting the algorithm to a random number generator or a secure pseudo-random number generator.

Following this line of thought, and using Bacteriophage T4 Genome (Miller *et al,* 2003) as a model, statistical tests for randomness were conducted on 1,000 numbers generated from the viral genome.

# METHODS

Bacteriophage T4 genome (Miller et al, 2003) was obtained from GenBank withm accession number AF158101. Complement 168,900' 167,900' (in the 5' to 3' direction), of T4 genome, representing 10,000 bases was investigated. The numbers were generated by using the numbers of the purine base (Adenine and Guanine) and pyrimidine bases (Thymine and Ctyosine) per turn of the DNA helix (10 bases) as digits.

For example, the first ten bases of the relate complement gave: Adenine, 4; Guanine, 1; Thymine3; Cytosine, 2. The next ten bases gave: Adenine, 0; Guanine, 2; Thymine, 5; Guanine, 2. Thus we get our first two numbers as 41,52. The numbers of the purine and pyrimidine bases were used alternatively per turns of the helix in order to preserve independence of the numbers, a quality of random numbers. The purine and pyrimidine numbers were not generated from the same helix (ten bases), as it would then be possible to predict something about the next number in the sequence from the preceding number, since all the digits would add up to ten, the number of bases per turn of the DNA helix (Brock & Madigan, 1991).

One thousand numbers were generated, which were subjected to three statistical tests for randomness: the Chi-squared test, the test of runs above and below the median, and the reverse arrangements test.

1. Summary statistics 2. Chi-squared test and Poisson distribution tables 3.Test of runs above and below the median 4. Reverse Arrangements Test.

# FINDINGS

The summary statistics table shows that the summation property of random numbers holds in the numbers generated from T4 phage genome. Taking into account that the minimum and maximum values generated from the viral genome are 0 and 82 respectively, the number of duplications and non-occurring values are within the range expected of random numbers.

The numbers failed the Chi-squared test but passed the other two tests. The Chi-squared test is a parametric test and as such makes an assumption about the distribution of the numbers. It measures how closely a set of numbers follows a uniform distribution.

The numbers generated do not follow a uniform distribution but approximate to a normal distribution. The distribution of the purine and pyrimidine bases, and amino acids in Bacteriophage T4 genome follows a Poisson distribution.

The third base of each codon plays a lesser role in specifying an amino acid than the first two, and in most cases codons which specify the same amino acid differ only at the third base (Nelson & Cox, 2000). From the 64 codons which specify amino acids, a core of 16 double bases can be made out. Since the numbers were generated from pairs of bases, there was an interest in determining the distribution of these double bases. Their distribution also follows a Poisson distribution.

The numbers passed the other two, non-parametric tests.

# DISCUSSION

There is enough evidence to show that the numbers generated from Bacteriophage T4 genome are random. Of the four properties of random numbers uniformity, duplication, summation and independence, the latter three is possessed by the numbers. Of the four properties, the properties of summation and duplication are more binding than those of uniformity and independence.

The sample size of 1,000 numbers may not hold much statistical weight, yet it highlights the existence of random processes within deoxyribonucleic acid (DNA). The restrictions involved in the Reverse Arrangements Test necessitated the use of 500 numbers in five different experiments.

A. **SUMMARY STATISTICS FOR NUMBERS GENERATED FROM T4 PHAGE GENOME.**

| Count | Mean | Median | Max | Min | P(Even) | Number of Duplications | Number of non-occurring values |
|---|---|---|---|---|---|---|---|
| 1000 | 33.8 | 36 | 82 | 0 | 0.496 | 40 | 40 |

## B. THE CHI-SQUARED TEST.

Observed and Expected Frequencies for the Chi-Squared Test

| Category | Observed Frequency (Oi) | Expected Frequency (Ei) |
|---|---|---|
| 0.100 | 31 | 111.1 |
| 0.200 | 107 | 111.1 |
| 0.300 | 240 | 111.1 |
| 0.400 | 241 | 111.1 |
| 0.500 | 205 | 111.1 |
| 0.600 | 113 | 111.1 |
| 0.700 | 47 | 111.1 |
| 0.800 | 13 | 111.1 |
| 0.900 | 3 | 111.1 |
| **Total** | **1000** | **999.9** |

H$_o$: The numbers follow a uniform distribution.
H$_A$: The numbers follow another distribution.

Test Statistic: $\chi^2 = \sum_{i=1}^{k} \frac{(O_i - E_i)}{E_i}$

$\chi^2 = 74.17$

Level of significance: α = 0.05

Critical value: $\chi^2_{0.05,\ 8} = 15.51$

The test statistic is more than the critical value so the null hypothesis is rejected at the 5% level of significance. The numbers do not follow a uniform distribution, one of the properties of random numbers. The numbers approximate to a normal distribution.

**TABLES SHOWING POISSON DISTRIBUTION OF PURINE BASES, PYRIMIDINE BASES AND AMINO ACIDS IN 3,500 BASES ( COMPLEMENT 168,900' – 165,400' IN THE 5' TO 3' DIRECTION IN BACTERIOPHAGE T4 GENOME.**

Adenine
µ=3.238

| X | P(x) | Theoretical frequency (fe) | Actual frequency |
|---|---|---|---|
| 0 | 0.0392 | 13.3 | 11 |
| 1 | 0.1269 | 43.2 | 35 |
| 2 | 0.2055 | 69.9 | 59 |
| 3 | 0.2218 | 75.4 | 92 |
| 4 | 0.1795 | 61.0 | 75 |
| 5 | 0.1162 | 39.5 | 41 |
| 6 | 0.0628 | 21.3 | 22 |
| 7 | 0.0290 | 9.7 | 5 |
|   | 0.8592 | 333.3 | 340 |

Thymine
µ=3.1706

| X | P(x) | Theoretical frequency(fe) | Actual frequency |
|---|---|---|---|
| 0 | 0.0419 | 14.2 | 9 |
| 1 | 0.1328 | 45.2 | 36 |
| 2 | 0.2106 | 71.6 | 81 |
| 3 | 0.2226 | 75.7 | 76 |
| 4 | 0.1764 | 59.9 | 68 |
| 5 | 0.1119 | 38.0 | 46 |
| 6 | 0.0591 | 20.1 | 18 |
| 7 | 0.0268 | 9.1 | 6 |
|   | 0.8768 | 333.8 | 340 |

Guanine
µ=1.5612

| X | P(x) | Theoretical frequency(fe) | Actual frequency |
|---|---|---|---|
| 0 | 0.2099 | 70.3 | 62 |
| 1 | 0.3277 | 109.8 | 110 |
| 2 | 0.2558 | 85.7 | 98 |
| 3 | 0.1331 | 44.6 | 46 |
| 4 | 0.0520 | 17.4 | 16 |
| 5 | 0.0162 | 5.4 | 3 |
|   | 0.9947 | 333.2 | 335 |

**Cytosine**
μ=2.0264

| X | P(x) | Theoretical frequency(fe) | Actual frequency |
|---|---|---|---|
| 0 | 0.1318 | 44.9 | 39 |
| 1 | 0.2671 | 91.1 | 78 |
| 2 | 0.2706 | 92.3 | 111 |
| 3 | 0.1828 | 62.3 | 74 |
| 4 | 0.0926 | 31.6 | 29 |
| 5 | 0.0375 | 12.8 | 8 |
| 6 | 0.0127 | 4.3 | 1 |
| 7 | 0.0037 | 1.3 | 1 |
|  | **0.9988** | **340.6** | **341** |

∗For the bases, A+T+G+C=3393. The actual number is 3500.The error arose from hand counting.

**Glycine**
μ=1.1020

| X | P(x) | Theoretical frequency(fe) | Actual frequency |
|---|---|---|---|
| 0 | 0.3322 | 16.3 | 19 |
| 1 | 0.3661 | 17.9 | 13 |
| 2 | 0.2017 | 9.9 | 11 |
| 3 | 0.0741 | 3.6 | 5 |
| 4 | 0.0204 | 1.0 | 1 |
|  | **0.9945** | **48.7** | **49** |

**Alanine**
μ=1.1224

| X | P(x) | Theoretical frequency(fe) | Actual frequency |
|---|---|---|---|
| 0 | 0.3255 | 15.9 | 17 |
| 1 | 0.3653 | 17.9 | 15 |
| 2 | 0.2050 | 10.0 | 13 |
| 3 | 0.0767 | 3.8 | 2 |
| 4 | 0.0215 | 1.1 | 2 |
|  | **0.9945** | **48.7** | **49** |

**Valine**
μ=1.8367

| X | P(x) | Theoretical frequency(fe) | Actual frequency |
|---|---|---|---|
| 0 | 0.1593 | 7.8 | 7 |
| 1 | 0.2925 | 14.3 | 16 |
| 2 | 0.2687 | 13.2 | 12 |
| 3 | 0.1645 | 8.1 | 9 |
| 4 | 0.0755 | 3.7 | 3 |
| 5 | 0.0277 | 1.4 | 1 |
| 6 | 0.0085 | 0.4 | 1 |
|   | **0.9967** | **48.9** | **49** |

**Leucine**
μ=1.4694

| X | P(x) | Theoretical frequency(fe) | Actual frequency |
|---|---|---|---|
| 0 | 0.2301 | 11.2 | 9 |
| 1 | 0.3381 | 16.6 | 14 |
| 2 | 0.2484 | 12.2 | 14 |
| 3 | 0.1217 | 5.96 | 8 |
| 4 | 0.0447 | 2.2 | 4 |
|   | **0.9830** | **48.16** | **49** |

**Isoleucine**
μ=1.4490

| X | P(x) | Theoretical frequency(fe) | Actual frequency |
|---|---|---|---|
| 0 | 0.2348 | 11.5 | 11 |
| 1 | 0.3402 | 16.7 | 19 |
| 2 | 0.2465 | 12.1 | 11 |
| 3 | 0.1191 | 5.8 | 5 |
| 4 | 0.0431 | 2.1 | 1 |
| 5 | 0.0125 | 0.6 | 1 |
| 6 | 0.0030 | 0.1 | 1 |
|   | **0.9992** | **48.9** | **49** |

**Methionine**
μ=0.6531

| X | P(x) | Theoretical frequency(fe) | Actual frequency |
|---|---|---|---|
| 0 | 0.5204 | 25.5 | 27 |
| 1 | 0.3399 | 16.7 | 14 |
| 2 | 0.1110 | 5.4 | 6 |
| 3 | 0.0242 | 1.2 | 2 |
|   | **0.9955** | **48.8** | **49** |

**Phenylalanine**
μ= 1.1429

| X | P(x) | Theoretical frequency(fe) | Actual frequency |
|---|---|---|---|
| 0 | 0.3189 | 15.6 | 18 |
| 1 | 0.3645 | 17.9 | 15 |
| 2 | 0.2083 | 10.2 | 8 |
| 3 | 0.0793 | 3.9 | 7 |
| 4 | 0.0227 | 1.1 | 1 |
|   | **0.9937** | **48.7** | **49** |

**Tyrosine**
μ=0.9184

| X | P(x) | Theoretical frequency(fe) | Actual frequency |
|---|---|---|---|
| 0 | 0.3992 | 19.6 | 17 |
| 1 | 0.3666 | 17.9 | 23 |
| 2 | 0.1684 | 8.2 | 5 |
| 3 | 0.0515 | 2.5 | 4 |
|   | **0.9857** | **48.3** | **49** |

**Tryptophan**
μ=0.3673

| X | P(x) | Theoretical frequency(fe) | Actual frequency |
|---|---|---|---|
| 0 | 0.6926 | 33.9 | 34 |
| 1 | 0.2544 | 12.5 | 12 |
| 2 | 0.0467 | 2.3 | 3 |
|   | **0.9937** | **48.7** | **49** |

**Serine**
μ=1.4082

| X | P(x) | Theoretical frequency(fe) | Actual frequency |
|---|---|---|---|
| 0 | 0.2446 | 11.9 | 12 |
| 1 | 0.3444 | 16.9 | 17 |
| 2 | 0.2425 | 11.9 | 11 |
| 3 | 0.1138 | 5.6 | 8 |
| 4 | 0.0401 | 1.9 | 0 |
| 5 | 0.0113 | 0.6 | 0 |
| 6 | 0.0026 | 0.1 | 1 |
|   | **0.9993** | **48.9** | **49** |

**Proline**
μ=0.6122

| X | P(x) | Theoretical frequency(fe) | Actual frequency |
|---|---|---|---|
| 0 | 0.5422 | 26.6 | 29 |
| 1 | 0.3319 | 16.3 | 10 |
| 2 | 0.1016 | 4.9 | 10 |
|   | **0.9757** | **47.8** | **49** |

**Threonine**
μ=1.0408

| X | P(x) | Theoretical frequency(fe) | Actual frequency |
|---|---|---|---|
| 0 | 0.3532 | 17.3 | 22 |
| 1 | 0.3676 | 18.0 | 12 |
| 2 | 0.1913 | 9.4 | 8 |
| 3 | 0.0664 | 3.3 | 5 |
| 4 | 0.0173 | 0.8 | 2 |
|   | 0.9926 | 48.8 | 49 |

**Cysteine**
μ=0.3469

| X | P(x) | Theoretical frequency(fe) | Actual frequency |
|---|---|---|---|
| 0 | 0.7069 | 34.6 | 32 |
| 1 | 0.2452 | 12.0 | 17 |
|   | **0.9521** | **46.6** | **49** |

**Asparagine**
µ=1.1633

| X | P(x) | Theoretical frequency(fe) | Actual frequency |
|---|---|---|---|
| 0 | 0.3125 | 15.3 | 12 |
| 1 | 0.3635 | 17.8 | 24 |
| 2 | 0.2114 | 10.4 | 8 |
| 3 | 0.0820 | 4.0 | 3 |
| 4 | 0.0238 | 1.2 | 2 |
|  | **0.9932** | **48.7** | **49** |

**Glutamate**
µ=0.7755

| X | P(x) | Theoretical frequency(fe) | Actual frequency |
|---|---|---|---|
| 0 | 0.4605 | 22.6 | 22 |
| 1 | 0.3571 | 17.5 | 18 |
| 2 | 0.1385 | 6.8 | 7 |
| 3 | 0.0360 | 1.8 | 2 |
|  | **0.9921** | **48.7** | **49** |

**Lysine**
µ=1.3061

| X | P(x) | Theoretical frequency(fe) | Actual frequency |
|---|---|---|---|
| 0 | 0.2709 | 13.3 | 12 |
| 1 | 0.3538 | 17.3 | 20 |
| 2 | 0.2311 | 11.3 | 10 |
| 3 | 0.1006 | 4.9 | 4 |
| 4 | 0.0328 | 1.6 | 3 |
|  | **0.9892** | **48.4** | **49** |

**Histidine**
µ=0.3673

| X | P(x) | Theoretical frequency(fe) | Actual frequency |
|---|---|---|---|
| 0 | 0.6926 | 33.9 | 33 |
| 1 | 0.2544 | 12.5 | 14 |
| 2 | 0.0469 | 2.3 | 2 |
|  | **1.0404** | **48.7** | **49** |

**Arginine**
μ=0.8367

| X | P(x) | Theoretical frequency(fe) | Actual frequency |
|---|---|---|---|
| 0 | 0.4331 | 21.2 | 23 |
| 1 | 0.3624 | 17.8 | 14 |
| 2 | 0.1516 | 7.4 | 9 |
| 3 | 0.0604 | 2.9 | 3 |
|   | **1.0075** | **49.3** | **49** |

**Aspartate**
μ=0.8980

| X | P(x) | Theoretical frequency(fe) | Actual frequency |
|---|---|---|---|
| 0 | 0.4074 | 19.9 | 22 |
| 1 | 0.3658 | 17.9 | 13 |
| 2 | 0.1643 | 8.0 | 11 |
| 3 | 0.0492 | 2.4 | 3 |
|   | **0.9867** | **48.3** | **49** |

**Glutamate**
μ=0.9796

| X | P(x) | Theoretical frequency(fe) | Actual Frequency |
|---|---|---|---|
| 0 | 0.3755 | 18.4 | 21 |
| 1 | 0.3678 | 18.0 | 13 |
| 2 | 0.1802 | 8.8 | 10 |
| 3 | 0.0588 | 2.9 | 5 |
|   | **0.9823** | **48.1** | **49** |

∗For the amino acids, the related complement is 168903'- 165585', in the 5' to 3' direction. The amino acids were counted in rows of 20.

# THE REVERSE ARRANGEMENTS TEST

Ho: The numbers generated do not exhibit monotonic trends
HA: The numbers generated exhibit monotonic trends

Test Statistic: $\sum_{j=1}^{N-1} \left[ \sum_{j=i+1}^{N} h_{ij} \right]$ where $h_{ij} = \begin{cases} 1 \text{ if } X_i > X_j \\ 0 \text{ else} \end{cases}$

$$A = \sum_{J=1}^{99} \left[ \sum_{j=i+1}^{100} h_{ij} \right]$$

A = 2308

Level of significance : α = 0.05

Critical Value : $A_N : (1 - \alpha/2) < 2308 \leq A_{N;(\alpha/2)}$

$A_{100;\,0.975} < 2308 \leq A_{100;\,0.025}$

$2145 < 2308 \leq 2804$

The value of A in this and the other four experiments lies between 2145 and 2804. The null hypothesis is accepted at the 5% level of significance.

**Test Values For the Five Experiments in the Reverse Arrangements Test.**

| A_1 | A_2 | A_3 | A_4 | A_5 | Avg A |
|------|------|------|------|------|-------|
| 2308 | 2262 | 2336 | 2267 | 2167 | 2268  |

**TEST OF RUNS ABOVE AND BELOW THE MEDIAN**

Summary Statistics for the Runs Test

| u | $n_1$ | $n_2$ | $\mu_u$ | $\sigma_u$ | z |
|---|---|---|---|---|---|
| 473 | 415 | 570 | 481.30 | 15.30 | - 0.51 |

Ho: The numbers are generated in a random order
HA: The numbers are not generated in a random order

The mean of the distribution of u :

$$\mu_u = \frac{2 n_1 n_2}{n_1 + n_2} + 1$$

The standard deviation of u, $\sigma_u$

$$= \sqrt{\frac{2 n_1 n_2 (2 n_1 n_2 - n_1 - n_2)}{(n_1 + n_2)^2 (n_1 + n_2 - 1)}}$$

The Test Statistic : $z = \frac{(u \pm 0.5) - \mu_u}{\sigma_u}$

$$z = - 0.51$$

Level of Significance : $\alpha = 0.05$

Critical Value : $/z/ < z_{\alpha/2}$

- $z - \alpha/2 < - 0.51 < z_{\alpha/2}$

- $z - _{0.025} < - 0.51 < z_{\ 0.025}$

- $1.96 < - 0.51 < +1.96$

The value of z lies between ± 1.96. The null hypothesis is accepted at the 5% level of significance.